\documentclass[11pt]{article} 
\usepackage[OT1]{fontenc}
\usepackage{authblk,amsthm,amsmath, amsfonts, caption, multirow, float, graphicx, tikz, pdfpages, setspace, enumitem, fancyhdr, longtable, accents} 
\usepackage[symbol]{footmisc}
\usepackage[misc]{ifsym}
\usepackage[square,numbers]{natbib}
\usetikzlibrary{arrows,shapes,shapes.geometric}

\usepackage[colorlinks,citecolor=blue,urlcolor=blue]{hyperref}
\usepackage[left=0.65in,right=0.65in,top=0.85in,bottom=0.85in,headsep=10mm,footskip=15mm]{geometry}

\usepackage{multicol}
\usepackage{xcolor}
\definecolor{green1}{rgb}{0.0, 0.5, 0.0}
\definecolor{chocolate}{rgb}{0.48, 0.25, 0.0}
\definecolor{dukeblue}{rgb}{0.0, 0.0, 0.61}
\colorlet{brown1}{brown!70!black}
\colorlet{blue1}{blue!70!black}
\colorlet{notgreen}{blue!50!yellow}
\usepackage{array}
\newcolumntype{L}[1]{>{\raggedright\let\newline\\\arraybackslash\hspace{0pt}}p{#1}}
\newcolumntype{X}[1]{>{\raggedright\let\newline\\\arraybackslash\hspace{0pt}}m{#1}}
\newcolumntype{C}[1]{>{\centering\let\newline\\\arraybackslash\hspace{0pt}}m{#1}}
\newcolumntype{R}[1]{>{\raggedleft\let\newline\\\arraybackslash\hspace{0pt}}m{#1}}

\usepackage{soul}
\usepackage[framemethod=tikz]{mdframed}

\mdfdefinestyle{myFigureBoxStyle}{backgroundcolor=black!5,, roundcorner=3pt,tikzsetting={draw=black, line width=1pt}}%

\makeatletter
\newcommand\fs@myRoundBox{\def\@fs@cfont{\bfseries}\let\@fs@capt\floatc@plain
	\def\@fs@pre{\begin{mdframed}[style=myFigureBoxStyle]}%
		\def\@fs@mid{\vspace{\abovecaptionskip}}%
		\def\@fs@post{\end{mdframed}}\let\@fs@iftopcapt\iffalse}
\makeatother

\theoremstyle{plain}

\newtheorem*{result*}{Result}

\theoremstyle{definition}

\newtheorem*{example*}{Example}

\newcommand{\eps}{\epsilon}

\newcommand{\vbar}{\, | \,}

\newcommand{\bS}{\ensuremath{\textbf{\emph{S}}}}

\newcommand{\z}{\ensuremath{\textbf{\emph{z}}}}

\newcommand{\y}{\ensuremath{\textbf{\emph{y}}}}
\newcommand{\Y}{\ensuremath{\textbf{\emph{Y}}}}

\newcommand{\ttheta}{\ensuremath{\boldsymbol{\theta}}}

\newcommand{\tbf}[1]{\noindent\textbf{#1}}

\title{\LARGE Estimating Software Reliability Using Size-biased Modelling}

\author[1]{Soumen Dey \footnote{\tbf{E-mail}: \href{mailto:soumenstat89@gmail.com}{\textit{soumenstat89@gmail.com}} \tbf{Orcid}:  \url{https://orcid.org/0000-0001-6270-2356} }}
\author[2]{Ashis Kumar Chakraborty \footnote{\tbf{Orcid}:  \url{https://orcid.org/0000-0002-2003-1336} }}

\affil[1]{Faculty of Environmental Sciences and Natural Resource Management, Norwegian University of Life Sciences, $\mathring{\text{A}}$s, Norway}

\affil[2]{Indian Statistical Institute, Kolkata, India}

\begin{document}
\tikzstyle{every picture}+=[remember picture] 
		
\everymath{\displaystyle}

\setlength{\abovedisplayskip}{5pt}
\setlength{\belowdisplayskip}{5pt}
\date{}
\maketitle




\begin{multicols}{2}
{\footnotesize
\begin{center}	{\bf Abstract } \end{center}

Software reliability estimation is one of the most active areas of research in software testing. Since time between failures (TBF) has often been challenging to record, software testing data are commonly recorded as test-case-wise in a discrete set up. 
We have developed a Bayesian generalised linear mixed model (GLMM) based on software testing detection data and a size-biased strategy which not only estimates the software reliability, but also estimates the total number of bugs present in the software. Our approach provides a flexible, unified modelling framework and can be adopted to various real-life situations. 
We have assessed the performance of our model via simulation study and found that each of the key parameters could be estimated with a satisfactory level of accuracy. We have also applied our model to two empirical software testing data sets. 
While there can be other fields of study for application of our model (e.g., hydrocarbon exploration), we anticipate that our novel modelling approach to estimate software reliability could be very useful for the users and can potentially be a key tool in the field of software reliability estimation.

}

	

{\footnotesize 
\tbf{Keywords}:  Software reliability, size-biased, Bayesian analysis, bug size, software testing
}


\section{Introduction}

Softwares are fuel of the current world. Economy, technology, transport, communication, medical treatment - all of these essential components of our daily lives are critically dependent on successful execution of softwares.  Most of the modern day devices may not function properly if the concerned softwares carry bugs. Thus, it is not surprising that estimation of software reliability remains a cornerstone in the field of software development and testing \citep{Pham2000Reliability, Shigeru2014Reliability}.



Determination of optimum time for software release remains an interesting field of research \citep{Chakraborty2019Bayesian}. It has also been proposed that the software testing data needs to be collected in a relatively efficient manner than the traditional process \citep{Nayak1988Estimation}. Time between failures (TBF) data have become difficult to collect as the complicacies in software development and its testing increases. In most cases, the logged information during software testing is test-case specific and consequently discrete in nature. Estimation of optimum duration of software testing under a discrete set up has also received considerable attention \citep{Chakraborty1994Optimum, Chakraborty1996Software, Dewanji2011Reliability, Das2017Optimum}. In these literature, optimum testing strategies have been developed based on the number of remaining bugs present in the software \citep{Chakraborty2019Bayesian, Heungseop2013Software}. 
However, if the remaining bugs are present in paths/locations (of the software) which will rarely be traversed by any inputs to be used by the users, the chances of software failure will also be rare, which in turn, could not solely infer the software to be unreliable. Though, this particular phenomenon seems plausible and close to reality, it has not been systematically studied yet in the literature.

In order to account for the probability of a remaining bug to be the cause of failure of a software, we introduce a latent variable, `eventual size of a bug' \citep{Chakraborty1996Software}. The eventual size of a bug is defined as the number of inputs that may eventually pass through the bug during the entire lifetime of a software, irrespective of whether the bug is detected or not during the testing phase. Occasionally, the eventual size of a bug is also referred to as simply `the size of the bug'. A software can be considered as a collection of several paths and each input to the software is expected to follow a particular path. In particular, if the same input is used several times, it can only check whether that particular path has any bugs or not, but it will not be able to check the presence of bugs in other paths as the given input will not traverse those other paths. A software would be require different inputs  to check existence of bugs in different paths. We may assume that an input can only identify at most one bug which lies on the path that the input would traverse. This size-biased approach was first introduced by \cite{Chakraborty1996Software} in software reliability, although the concept had also been applied in a few other fields of investigation \citep{Patil1978sizebiased}.
 

\subsection{Some terminologies in software testing} \label{sec:size.biased}

\tbf{Differential sizes of the bugs in paths, sub-paths of a software.} It is quite natural that a path in the software branches into several sub-paths at a later stage. For all these sub-paths, a part of the paths is common for all in the beginning. Now suppose that a bug is present on the common path and another bug is on one of the several sub-paths associated with the common path. It is quite natural that the size of the bug in the common path is much higher compared to that of the bug in the sub-paths, since the inputs passing through each of the sub-paths must have traversed through the common path before entering into a sub-path. The size of a bug also, thus, may give an indication of how quickly a bug could be identified. If a bug of large size is not detected, it may be a potential threat to the functioning of the software. Thus, it is straightforward that the probability of detection of a bug depends on the size of the bug. Larger the size of the bug, larger will be the probability of detecting that bug, as has been indicated in \cite{Chakraborty1994Optimum}. In fact, they have also shown that similar concepts are applicable in discovering fields with rich hydrocarbon contents in the field of producing oil and natural gas. 

\tbf{Software reliability and its dependence on location of a bug.} A bug which exists in a path that is rarely traversed by any input, is likely to be harmless as far as running of the software is concerned. Thus, reliability of the software does not only depend on the number of bugs remaining in the software, it also depends on the positioning of the bugs, particularly the paths on which it exists and whether that path is frequently traversed by random inputs from the user \citep{Littlewood1979Reliability}. Hence in order to have a better model for software reliability, our attention would be to find out the total size of the bugs that will remain and not just the number of remaining bugs.

\tbf{Software testing and different testing phases.} In a discrete software testing framework, when an input is being tested, it results in either a success (i.e., finding an error) or a failure (i.e., not finding an error). Testing of software is carried out into many phases, where, in each phase a series of inputs are tested and results of each testing are recorded as either a success or a failure \citep{Dewanji2011Reliability}. After identifying the bugs at the end of testing within a phase, they are debugged at the end of the phase. This process of debugging is known as periodic debugging or interval debugging \citep{Das2016Reliability}.

Hence, detecting a bug during software testing, can be thought to be a probabilistic sampling, where the chances of a bug being detected is an increasing function of the size of the bug. This is analogous to the size-biased modeling by \cite{Patil1978sizebiased}, for modeling identification of species.

\subsection{Motivation}
The present work was motivated by one key idea from one of the authors in early nineties, that the optimum time to stop software testing and the optimum time to stop drilling in hydrocarbon exploration, were found to be analogous \citep{Chakraborty1994Optimum, Chakraborty1996Software}. It is quite logical to understand that bigger field, in terms of the amount of oil and natural gas that can be obtained after drilling in, are expected to be drilled much ahead compared to the others. Thus, size (in terms of the value of the oil and natural gas in the field) plays an important role in identifying the chronological order of the drilling areas. Ideally, this strategy would minimize the overall drilling cost. Similarly, once the size of a bug is appropriately defined (as has been done in earlier paragraphs), the size-biased nature of the problem can be used to model the detection probability of a bug in software testing data. However, the eventual size of a bug remains unknown, which becomes the major challenge to the present problem.

The article is organized as follows. In Section~\ref{methods}, we developed a Bayesian generalized linear mixed model, whereas in Section~\ref{sec:model.fitting}, we provided a description of model fitting and model performance measures used for this study. In Section~\ref{sec:simstudy}, we showcased a simulation study to assess the performance of our model. We assessed the performance of the models using relative bias, coefficient of variation, and coverage probability. Application of this model was carried out on two empirical software testing data sets. Section~\ref{sec:application.software.bugdata} illustrates an application of the developed Bayesian model to a commercial software testing data set and in Section~\ref{sec:application.ISRO.data}, we showed an application of the model to a very critical software data set used for space mission software testing. The article ends with a discussion and conclusion in Section~\ref{sec:discussion}. 

\section{Methods}\label{methods}
\subsection{General approach}\label{sec:gen.approach}
We utilized the hierarchical modelling philosophy to formulate a statistical model to address the problem of estimating the total number of bugs in the presence of imperfect detection of the bugs during software testing process. The developed model can also be used to estimate the remaining eventual bug size that are present in the software, as well as the software reliability. We also provided a new method to predict the stopping phase such that the estimated remaining bug size at that phase remains below a preassigned threshold. Later, we extended the model described above to also accommodate the possible groups of bugs who share the same bug size.

\subsection{Model description}\label{sec:bugmodel}

The model has composed of two hierarchical structure: one for the state process that explains the latent dynamic of the bugs within the software, while the other part corresponds to the observation model explaining the probabilistic structure of the observed software testing data.

\subsubsection{State process}\label{stateprocess}
Consider $N$ number of distinct and independent bugs are present in a particular software and size of each bug is denoted by $S_i$, $i= 1,2,\dots,N$. The eventual size of a bug (or in short, size of a bug) is considered as a latent variable in the model and is needed to be estimated. 
Let $\bS$ denotes a vector of these latent variables $S_1, S_2, \dots, S_N$ defining the size of the $N$ (unknown) bugs under study. For the ease of computation and other technical advantages (described later), we define $N \sim \mathrm{Binomial}\left(M,\psi\right)$, where $M$ represents the maximum possible number of bugs present in the software and $\psi$ denotes the inclusion probability to indicate the proportion of $M$ that represent the real population of bugs. 


\subsubsection{Observation process}\label{observationprocess}
We suppose that $T_j, \, j = 1,2,  \dots, Q$ inputs are used for each of the $Q$ testing phases. We consider the situation where a present bug can get detected in any of the $T_j$ inputs at the $j$-th phase, $j = 1,2,  \dots, Q$. 

Let $y_{ij}$ represent the binomial detection outcome for a bug $i$ over the $T_j$ inputs at phase $j$. If $y_{ij} > 0$, this subsequently implies that $y_{il} = 0$, $l = 1, 2, \dots, (j-1)$. 
It should be noted that after a bug gets detected at the $j$-th phase, it is eliminated from the pool of bugs during the debugging at the end of phase $j$. For example, in a software testing, if bug 1 gets detected at phase $j=4$, we would have $y_{11} = y_{12} = y_{13} = 0$ and $y_{14} > 0$. 

We used the data augmentation approach to model the number $N$ of bugs in the software by choosing a large integer $M$ to bound $N$ and introduced a vector of $M$ latent binary variables $\z = (z_1,z_2,\dots,z_M)$ such that $z_i = 1$ if individual $i$ is a member of the population and $z_i = 0$ otherwise. We assume that each $z_i$ is a realisation of a Bernoulli trial with parameter $\psi$, the inclusion probability.

A binomial model, conditional on $z_i$, is assumed for each observation $y_{ij}$:
\begin{align}
y_{ij} \sim \text{Binomial}(T_j, p_{i}z_{i}), 
\end{align}
where $p_{i}$ denotes the detection probability of the $i$-th bug in a phase. The detection probability $p_{i}$ is modelled as a increasing function of the bug size $S_{i}$, since the detection probability directly depends on the size of a bug, that is, more the bug size, higher the detectability. 

\subsubsection{Model for detection probability}\label{detprobmodel}
From the definition of bug size, $S_i$ is higher if placement of $i$-th bug is on a common path near the origin and a number of sub-paths follow subsequently. If $r$ denotes the probability of bug detection in any one of the inputs that will pass through the $i$-th bug, then the probability of detecting $i$-th bug with one input is
\begin{align}
p_{i} =  p(r, S_{i}) = 1- (1-r)^{S_i}.
\end{align}
The parameter $r$ plays the role of a shared parameter across all the bugs and critical for the dependence structure of the nodes in our joint probability model. In addition, the above formulation of $p_{i}$ comes naturally from our definition of bug size and accounts for individual-level heterogeneity in detection probability of the bugs \citep{Patil1978sizebiased}. Note that, here $p_{i}$ is modelled as a monotonically increasing function of $S_{i} $ and when $ S_{i} = 0$, we have $p_{i} = 0$.

\subsubsection{Model for \emph{N}}\label{Nmodel}
We used the data augmentation approach to model the number of bugs $N$ in the software by choosing a large integer $M$ to bound $N$ and introduced a vector of $M$ latent binary variables $\z = (z_1,z_2,\dots,z_M)$ such that $z_i = 1$ if individual $i$ is a member of the population and $z_i = 0$ otherwise. We assume that each $z_i$ is a realisation of a Bernoulli trial with parameter $\psi$, the inclusion probability.

We assume that $n$ bugs get detected over the $Q$ testing phases which is expected to be less than the total number of bugs $N$ due to imperfect detection during testing. Consequently, as part of the data augmentation approach, the detection data set  $\{y_{ij}\}_{i,j}$ is supplemented with a large number of ``all-zero'' encounter histories $\Y_{\text{rem}}$, an array of ``all-zero'' detection histories with dimensions $(M-n) \times Q$. We label the zero augmented complete detection data set as $\Y$.

\subsubsection{Estimating the remaining eventual bug size and the stopping phase}\label{sec:stoppingphase}

In software testing, certain decisions are critical: for example, when should we stop testing, what should be the criteria to stop software testing process. If after the testing and debugging phases, certain bugs remain in the software, it may cause improper functioning of the software even after the market release. Therefore, a decision to optimize software testing and debugging time is an important part of the development process of software. 

The above model is well suited to estimate the number of bugs $N$, the detection probability $p_i$'s, and bug size $S_i$'s. But to estimate the remaining eventual bug size at a later untested phase, we proceed as follows. 

We denote $ f $ as the model for the detection observations for a bug with number of inputs $T_j$, $j = 1,2, \dots, J$, where $J > Q$, and $ \tilde{\y} $ as future
observation or alternative detection outcome that could have been obtained during the testing phase. Since the stopping phase (such that the remaining eventual total size of the bugs is less than a threshold, say, $\epsilon$) is unknown to the software tester, we assign a sufficiently large value for $J$, considering the available RAM size of the computing device and  and computing time. The
posterior predictive model for a new detection data $ \tilde{\y}_i $ for the $i$-th bug is then,
\begin{align}
f(\tilde{\y}_i \vbar \Y) = \int f(\tilde{\y}_i \vbar \ttheta) \pi(\ttheta \vbar \Y) \, d\ttheta
\end{align}
where $ \ttheta $ denotes the vector of all the parameters $ r, \bS, \z, \psi $ and $ f(\tilde{\y}_i \vbar \Y)  $ is the predictive density for $ \tilde{\y}_i $ induced by the posterior distribution $ \pi(\ttheta \vbar \Y) $. 

In practice, we obtain a single posterior replicate $ \tilde{\y}_i^{(l)} $ by drawing from the model $ f(\tilde{\y}_i \vbar \ttheta^{(l)}) $, where $\{\ttheta^{(l)} \, : \, l = 1, 2, \dots, L\}$ represents a set of MCMC draws from the posterior distribution of parameter $\ttheta$. 

We define a set of deterministic binary variables $u_{ij}$ which takes the value 1 if $i$-th bug is detected on or before $j$-th phase and 0 otherwise. Total size of the bugs that are detected up to the $j$-th phase
is then computed as $A_j = \sum_{i = 1}^M S_i z_i u_{ij}$, $j = 1,2, \dots, J$. Consequently, we also compute the total eventual remaining size of the bugs that are not detected up to the $j$-th phase, $B_j = \sum_{i = 1}^M S_i z_i (1-u_{ij})$, $j = 1,2, \dots, J$. We obtain the stopping phase, denoted by $k$, such that $B_k < \eps$ (where $\eps$ is a preassigned threshold). We compute $B_j$ for each replicated data set $ \{\tilde{\y}_i^{(l)} \, : \, i=1,2, \dots, M\}$, $l=1,2, \dots, L$, thus enabling us to obtain an MCMC sample for both $k$ and $\{B_j \, : \, j = 1,2, \dots, J\}$.

\subsubsection{Software reliability}\label{sec:softwarereliability}
For software testing detection data set, we define software reliability, at a testing phase $j$, as the posterior probability that the total eventual remaining size of the bugs $B_j$ (that are not detected up to the $j$-th phase), is less than or equal to the prefixed small quantity $\eps$ given the observed detection data $\Y$, 
\begin{align}\label{reliabilityeq}
\gamma_j (\eps) = P(B_j \leq \eps \, | \, \Y). 
\end{align}
Consequently, reliability is a non-decreasing function of threshold $\eps$ and testing phase $j$. Asymptotically, for a fixed $j$, (i) as $\eps \rightarrow 0$, $\gamma_j (\eps) \rightarrow 0$ and (ii) as $\eps \rightarrow \infty$, $\gamma_j (\eps) \rightarrow 1$. Similarly, for a fixed $\eps$, if we conduct a very large number of testing phases (i.e., $j$ becomes large), reliability $\gamma_j (\eps)$ will be very close to 1. Of course, this rate of convergence will also depend on the number of testing inputs $T_j$ in each phase.

\subsection{Modelling for grouped bugs}\label{sec:groupmodel}
Often we come across situations where a few bugs are collocated on the same path or same part of the software in such a way that we can assume without loss of generality that each of them have the same bug size. 
 For computational and notational simplicity, we make a transformation of the data set $((y_{ij}))$ to $(y_g^*)$ where the observed data $y_{g}^*$ represents the number of bugs from the $g$-th group that are detected. Consequently, we have $y_{g}^* \sim \text{Binomial} (T_{j(g)}, p_g^*)$, $p_g^*$ denotes the probability of detecting a bug belonging to the $g$-th bug-group with a single test case and $j(g)$  denotes the corresponding phase to the $g$-th group.

Here, we consider a number of distinct group of bugs $N_G$ that are present in a software and each bug in a group (say, $g$-th) has size $S_g^*$. Each group of bugs comprises at least one bug. Following Section~\ref{stateprocess}, we define $N_G \sim \mathrm{Binomial}\left(M_G,\psi\right)$, where $M_G$ is a large positive integer that gives an upper bound to $N_G$.
The link between $p_g$ and the size $S_g^*$ remains the same as in Section~\ref{detprobmodel}, $p_{g}^* =  1- (1-r^*)^{S_g^*}$. We used the data augmentation approach to model the number of bug-groups $N^*$  (discussed in Section~\ref{Nmodel}). The total number of bugs $N^*$ has the following expression:
\begin{align}\label{Nstar}
N^* = n + \sum_{g = 1}^{M_G} a_g z_g
\end{align}
where $n$ denotes the number of bugs detected during the testing period and $a_g$ denotes the number of bugs in the $g$-th group that went undetected. We utilized the posterior predictive distribution of new detection data $\tilde{\y}^* $ with density $f(\tilde{\y}_g^* \vbar \Y^*)$ to estimate $a_g$.

To compute the remaining eventual size, we introduce binary variables $((u_{gQ}))$, $g = 1, 2, \dots, M_G$, where $u_{gQ}$ takes the value 1 if $g$-th bug-group is detected on or before $Q$-th phase and takes 0 otherwise. The remaining eventual size is calculated as $B_Q = \sum_{g = 1}^{M_G} S_g z_g d_g (1-u_{gQ})$, where $d_g$ denotes the number of bugs in $g$-th bug-group .

\subsection{Prior assignment}\label{sec:prior}
Bug sizes ($S_i$'s) are usually latent and unobservable. We assign a Poisson-Gamma mixture prior for $S_i$ to capture the required level of variability in the latent variable. Consequently, each $S_i$ is assumed to follow Poisson distribution with mean $\lambda_i$, where the $\lambda_i$ is a random draw from Gamma distribution with shape parameter $a_s$ and rate $b_s$. We assign bounded Uniform prior over the interval $(0,1)$ for detection probability $r$ and the inclusion probability $\psi$. These proper prior specifications ensured propriety of the posteriors. 


\section{Model fitting}\label{sec:model.fitting}

 We fitted models using Markov chain Monte Carlo (MCMC) simulations. In particular, we used Gibbs sampling for simulating the parameters from the posterior distribution. The full posterior of each $z_i$ is Bernoulli distributed random variables, whereas the full posteriors of the other parameters and latent variables (e.g., $\psi$, $r$, $S_i$'s) are of non-standard forms. We used slice sampler for $S_i$'s and random walk Metropolis-Hastings sampler for the other parameters (e.g., $\psi$, $r$). We implemented MCMC computations using NIMBLE \citep{valpine2017nimble} in R software  \citep{rsoftware}. We ran three chains of 10000 iterations including an initial burn-in phase of 5000 iterations. MCMC convergence and mixing of each model parameters was monitored using the Gelman-Rubin convergence diagnostics $\hat{R}$ \citep[with upper threshold 1.1]{gelman2014bayesian} and MCMC traceplots. 


\subsection{Model performance measures}\label{sec:model.performance}
We used relative bias, coefficient of variation and coverage probability to evaluate the effect of detection function misspecifications on population size and home range size estimators. Suppose $\{\theta^{(r)} \, : \, r = 1, 2, \dots, R\}$ denotes a set of MCMC draws from the posterior distribution of a scalar parameter $\theta$. 

\tbf{Relative bias}.
	Relative bias (RB) is calculated as 
\begin{align}
	\widehat{\text{RB}} (\theta) = \frac{\hat{\theta} - \theta_0}{\theta_0},
	\end{align}
	where $\hat{\theta}$ denotes the posterior mean $\frac{1}{R} \sum_{r=1}^{R} \theta^{(r)}$ and $\theta_0$ gives the true value.

\tbf{Coefficient of variation}.
	Precision was measured by the coefficient of variation (CV):
\begin{align}
	\widehat{\text{CV}} (\theta) = \frac{\widehat{\text{SD}}(\theta)}{\hat{\theta}},
	\end{align}
	where $\widehat{\text{SD}}(\theta) = \sqrt{\frac{1}{R} \sum_{r=1}^{R} (\theta^{(r)} - \hat{\theta})^2}$ is the posterior standard deviation of parameter $\theta$.

\tbf{Coverage probability}.
Coverage probability was computed as the proportion of model fits for which the estimated 95\% credible interval of the estimate (CI) contained the true value of $\theta$.



\section{Simulation study}\label{sec:simstudy}
  

\subsection{Description of simulated data and simulation scenarios}\label{sec:simdata}

For a complex high-dimensional model such as described in Section~\ref{sec:bugmodel}, it would be instrumental to assess model performance with respect to different ranges of the model parameters. We simulated detection data sets of software testing for two values of detection parameter $r$, viz., $0.75 \times 10^{-5}$ and $1.5 \times 10^{-5}$, and two values of number of inputs in each phase ($T_j$), viz., 1000 and 2000. In total we have four different simulation scenarios (viz., Sets 1-4) and we simulated a total of 200 data sets (i.e., 50 data sets under each scenario). In each scenario, we assumed a fixed number of bugs $N=200$ for simulating the detection data of bugs and the software testing was carried out over $Q=5$ phases. The key details of the simulated data sets are given in Table~\ref{table:ndets}. The number of detected bugs (and also the total number of detections) are higher on average (mean 132) in the set 2 with number of inputs as 2000 as compared to set 1 (mean 106) with number of inputs as 1000, detection parameter $r$ remains unchanged in both these two sets at $0.75 \times 10^{-5}$. Same phenomenon can be observed for sets 3 (number of inputs = 1000) and 4 (number of inputs = 2000) where $r = 1.5 \times 10^{-5}$ (see Figure~\ref{fig:sim.results}a,c). For estimating the remaining eventual bug size and the stopping phase, the posterior predictive simulations are carried out for 25 additional phases, implying $J=Q+25=30$ (see Section~\ref{sec:stoppingphase}).

\subsection{Results from Simulation study}\label{sec:simesults}
\begin{figure*}[ht] 
	\centering
	{\footnotesize \begin{tabular}{@{\hspace{-0.1cm}} l } 
			\includegraphics[scale=0.19]{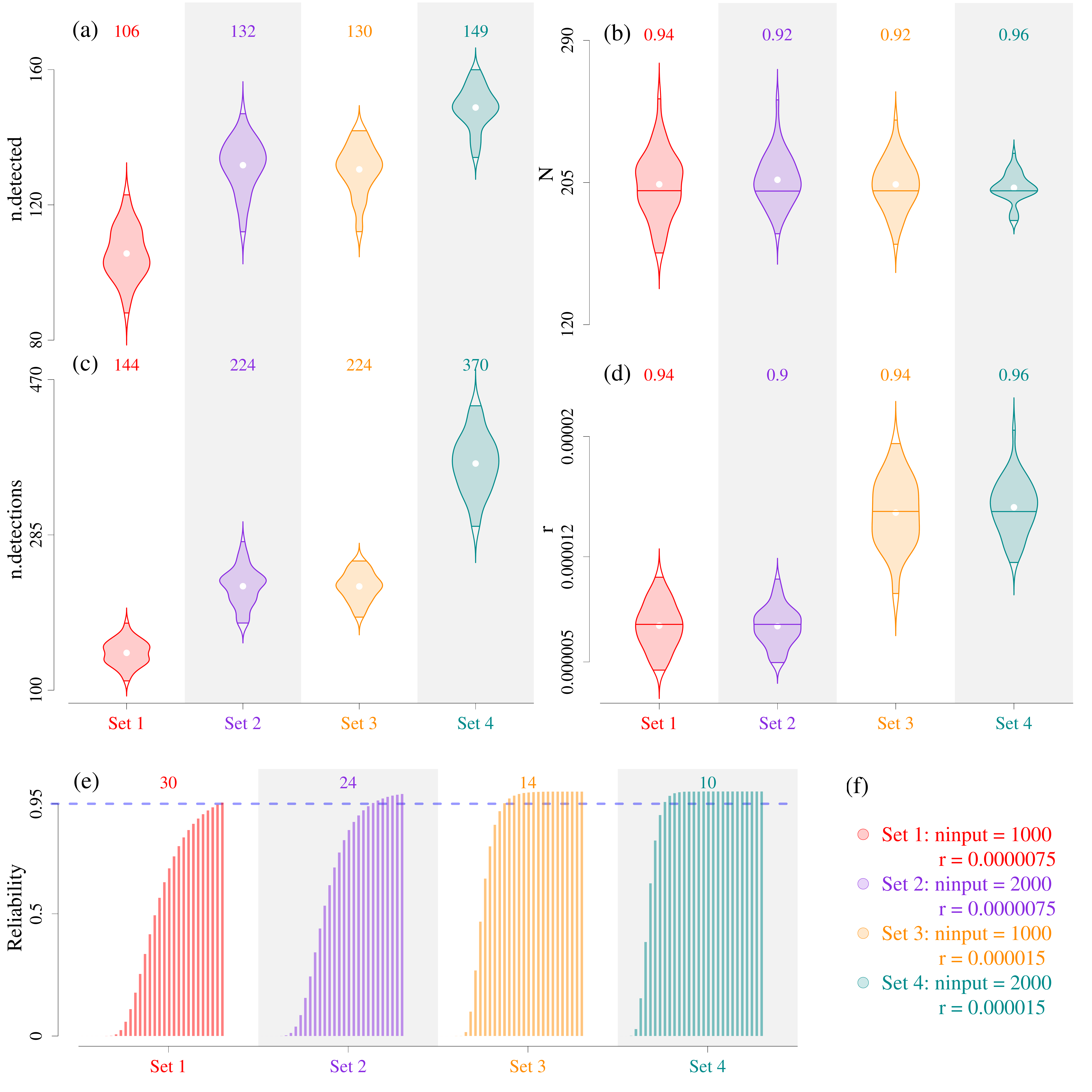}
	\end{tabular}}
	\caption{\footnotesize Details of simulated data summary and parameter estimates across the 4 simulation scenarios. Panels (a) and (c): Violins of the number of detected bugs (panel a) and the total number of detections across all the bugs, i.e., $\sum_{i=1}^M\sum_{j=1}^J y_{ij}$ (panel c) from the 50 simulated data sets in each of the 4 scenarios. The mean of the corresponding variable  in each panel (a) and (c) are mentioned at the top of each violin. Panels (b) and (d): Violins of posterior mean estimates of the population size estimator $N$ (panel b) and detection parameter $r$ (panel d). The violins represent the distribution over 50 simulated data sets of each simulation scenarios. Panel (e): Estimates of posterior reliability with threshold 100 and stopping phase for attaining optimum reliability level 0.95. The bars represent posterior reliability at phases 1,2,\dots,30. Panel (f): Pre-specified values of the number of inputs and $r$ in each set of simulation scenario that were used to simulate the data sets.}
	\label{fig:sim.results}
\end{figure*}

We fitted our Bayesian size-biased model to each of the 200 simulated data sets using MCMC and $M$ is set to 400 for each model fitting. All MCMC samples of the parameters of interest (e.g., population size $N$, detection parameter $r$) were obtained after ensuring proper mixing and convergence, with $\hat{R}$ values below 1.1. The posterior estimates of different parameters were obtained using the MCMC chains. The posterior summaries of the total number of bugs $N$ and detection parameter $r$ for the simulation study are provided in Table~\ref{table:N.est}, respectively and also portrayed in Figure~\ref{fig:sim.results}.

The relative bias and coefficient of variation of $N$ and $r$ are estimated for each of the 50 replicates in each set. The relative bias estimates of $N$ in each set varied between: (-16\%, 19\%) in set 1, (-9\%, 19\%) in set 2, (-12\%, 15\%) in set 3, (-9\%, 9\%) in set 4) and the coefficient of variation of $N$ in each set varied between: (8\%, 12\%) in set 1, (5\%, 7\%) in set 2, (6\%, 7\%) in set 3, (4\%, 5\%) in set 4. The relative bias estimates of $r$ in each set varied between: (-36\%, 37\%) in set 1, (-32\%, 32\%) in set 2, (-33\%, 27\%) in set 3), (-18\%, 22\%) in set 4 and the coefficient of variation of $r$ in each set varied between: (16\%, 24\%) in set 1, (13\%, 17\%) in set 2, (13\%, 17\%) in set 3, (11\%, 15\%) in set 4. Coverage probabilities of both $N$ and $r$ were higher than 90\% in each of the scenarios (Figure~\ref{fig:sim.results}).

We estimated the reliability at the end of each phase and also at different possible future phases (assuming a pre-specified  number of test cases in each phases). It is important to mention that, the estimation of reliability heavily depends on the pre-specified threshold and the number of test cases used during the future phases (that would be conducted after the first 5 phases already conducted). Here we have assumed that the number of test cases in each future phase to be the same as the number of inputs in the respective scenario.

The reliability (i.e., posterior probability of the remaining size lying below a threshold) is a non-decreasing function of testing phase index, since remaining bug size  gets reduced with more bugs being detected in subsequent testing phases. We found the reliability estimates to attain the targeted 95\% level (with threshold 100) to be varying with respect to different simulation scenarios (Figure~\ref{fig:sim.results}). For instance, the reliability estimate attained the optimum 95\% level (with threshold 100) at phase 30 in set 1, implying the developer would need to continue software testing for 25 more future phases (after the 5 testing phases already conducted) to attain optimum software reliability level. Hence, the stopping phase was estimated as 30. For other sets, the estimates of the stopping phases were at phase 24 (set 2), phase 14 (set 3) and phase 10 (set 4).

\end{multicols}

\begin{table*}[ht] 
\centering 
	\caption{\footnotesize Number of detected individuals and number of total detections (mean, median, 2.5\% and 97.5\% quantiles) in simulated SCR data sets across 50 repetitions for each simulation scenario.} 
	\label{table:ndets} 
	{\scriptsize\begin{tabular}{@{\hspace{0em}} l @{\extracolsep{8pt}}  l @{\extracolsep{8pt}}  r @{\extracolsep{5pt}}  r @{\extracolsep{5pt}}  r @{\extracolsep{5pt}}  r @{\extracolsep{5pt}} r @{\extracolsep{20pt}} r @{\extracolsep{5pt}} r @{\extracolsep{5pt}} r @{\extracolsep{5pt}} r} 
		\\[-1.8ex]\hline 
		\hline \\[-1.8ex] 
		\multirow{2}{*}{\hspace{-5pt}\begin{tabular}{l} Sl.\\no. \end{tabular}} & \multirow{2}{*}{\begin{tabular}{l} No. of\\inputs \end{tabular}\hspace{-5pt}} &
		\multirow{2}{*}{\begin{tabular}{l} $r$\hspace*{12pt}\\ \end{tabular}\hspace{-5pt}} &
		\multicolumn{4}{c}{No. of detected individuals} & \multicolumn{4}{c}{No. of detections} \\ [0.3em]
		\cline{4-7} \cline{8-11} \\[-0.5em]
		 & & & Mean & Median & 2.5\% & 97.5\% & Mean & Median & 2.5\% & 97.5\%  \\ 
		& & &  &  & Quantile & Quantile & &  & Quantile & Quantile  \\ 
		[0.5em] \hline \\[-1ex]  
$1$ & $1000$ & $0.75 \times 10^{-5}$ & $106$ & $104$ & $92$ & $121$ & $144$ & $145$ & $117$ & $171$ \\ 
$2$ & $2000$ & $0.75 \times 10^{-5}$ & $132$ & $133$ & $114$ & $144$ & $224$ & $228$ & $182$ & $266$ \\ 
$3$ & $1000$ & $1.5 \times 10^{-5}$ & $130$ & $131$ & $113$ & $142$ & $224$ & $224$ & $191$ & $253$ \\ 
$4$ & $2000$ & $1.5 \times 10^{-5}$ & $149$ & $149$ & $135$ & $159$ & $370$ & $370$ & $299$ & $432$ \\  [0.5em]

	\hline\\[-0.2em] 

	\end{tabular} }
\end{table*} 
\begin{table*}[ht] \centering 
	\caption{\footnotesize Relative bias (mean, median, 2.5\% and 97.5\% quantiles), coefficient of variation (mean, median, 2.5\% and 97.5\% quantiles) and coverage probability of the 95\% credible interval for population size $N$ and detection probability $r$ across 50 repetitions for each simulation scenario.}
	\label{table:N.est} 
{\scriptsize\begin{tabular}{@{\hspace{0em}} l @{\extracolsep{8pt}}  l @{\extracolsep{8pt}}  r @{\extracolsep{5pt}}  r @{\extracolsep{5pt}}  r @{\extracolsep{5pt}}  r @{\extracolsep{5pt}} r @{\extracolsep{20pt}} r @{\extracolsep{5pt}} r @{\extracolsep{5pt}} r @{\extracolsep{5pt}} r  @{\extracolsep{5pt}} r} 
		\\[-1.8ex]\hline 
		\hline \\[-1.8ex] 
		\multirow{2}{*}{\hspace{-5pt}\begin{tabular}{l} Sl.\\no. \end{tabular}} & \multirow{2}{*}{\begin{tabular}{l} No. of\\inputs \end{tabular}\hspace{-5pt}} &
		\multirow{2}{*}{\begin{tabular}{l} $r$\hspace*{12pt}\\ \end{tabular}\hspace{-5pt}} &
		\multicolumn{4}{c}{Relative bias} & \multicolumn{4}{c}{Coefficient of variation}&
		\multirow{2}{*}{\hspace{5pt}\begin{tabular}{r} Coverage\\ probability \end{tabular}\hspace{-10pt}} \\ [0.3em]
		\cline{4-7} \cline{8-11} \\[-0.5em]
		 & & & Mean & Median & 2.5\% & 97.5\% & Mean & Median & 2.5\% & 97.5\%  \\ 
		& & &  &  & Quantile & Quantile & &  & Quantile & Quantile & \\ 
		[0.5em] \hline \\[-1ex]  

	\multicolumn{12}{l}{\hspace{-5pt}Population size $N$ } \\[0.5em] 
$1$ & $1000$ & $0.75 \times 10^{-5}$ & $0.020$ & $0.020$ & $-0.160$ & $0.190$ & $0.100$ & $0.090$ & $0.080$ & $0.120$ & $0.940$ \\ 
$2$ & $2000$ & $0.75 \times 10^{-5}$ & $0.030$ & $0.030$ & $-0.090$ & $0.190$ & $0.060$ & $0.060$ & $0.050$ & $0.070$ & $0.920$ \\ 
$3$ & $1000$ & $1.5 \times 10^{-5}$ & $0.020$ & $0.020$ & $-0.120$ & $0.150$ & $0.060$ & $0.060$ & $0.060$ & $0.070$ & $0.920$ \\ 
$4$ & $2000$ & $1.5 \times 10^{-5}$ & $0.010$ & $0.010$ & $-0.090$ & $0.090$ & $0.050$ & $0.050$ & $0.040$ & $0.050$ & $0.960$ \\ [1em]

	\multicolumn{12}{l}{\hspace{-5pt}Detection probability $r$} \\[0.5em]  
$1$ & $1000$ & $0.75 \times 10^{-5}$ & $-0.010$ & $-0.010$ & $-0.360$ & $0.370$ & $0.190$ & $0.190$ & $0.160$ & $0.240$ & $0.940$ \\ 
$2$ & $2000$ & $0.75 \times 10^{-5}$ & $-0.020$ & $-0.010$ & $-0.320$ & $0.320$ & $0.150$ & $0.150$ & $0.130$ & $0.170$ & $0.900$ \\ 
$3$ & $1000$ & $1.5 \times 10^{-5}$ & $0$ & $-0.010$ & $-0.330$ & $0.270$ & $0.150$ & $0.150$ & $0.130$ & $0.170$ & $0.940$ \\ 
$4$ & $2000$ & $1.5 \times 10^{-5}$ & $0.020$ & $0.020$ & $-0.180$ & $0.220$ & $0.130$ & $0.130$ & $0.110$ & $0.150$ & $0.960$ \\  [0.5em]


	\hline\\[-0.2em] 

	\end{tabular} }

\end{table*}



\vspace{2cm}

\begin{multicols}{2}

\section{Application to commercial software testing empirical data}\label{sec:application.software.bugdata}

\subsection{Data description}\label{sec:software.bug.data}

The data set consists a total of 8757 test inputs detailed with build number, case id, severity, cycle, result of
test, defect id etc. In this data, the severity of a path is broadly divided into three categories, namely,
simple, medium and complex depending on the effect of the bug if it is not debugged before marketing
the software. The data has four cycles namely Cycle 1, Cycle 2, Cycle 3 and Cycle 4, which is equivalent
to the different phases of testing we have referred to Section~\ref{methods}. After each cycle, the bugs that are
identified during the cycle are debugged as mentioned in the Section~\ref{methods}. 





\subsection{Results from commercial software testing data analysis}\label{sec:software.results}
\begin{figure*}[ht] 
	\centering
	{\footnotesize \begin{tabular}{@{\hspace{-0.1cm}} l } 
			\includegraphics[scale=0.19]{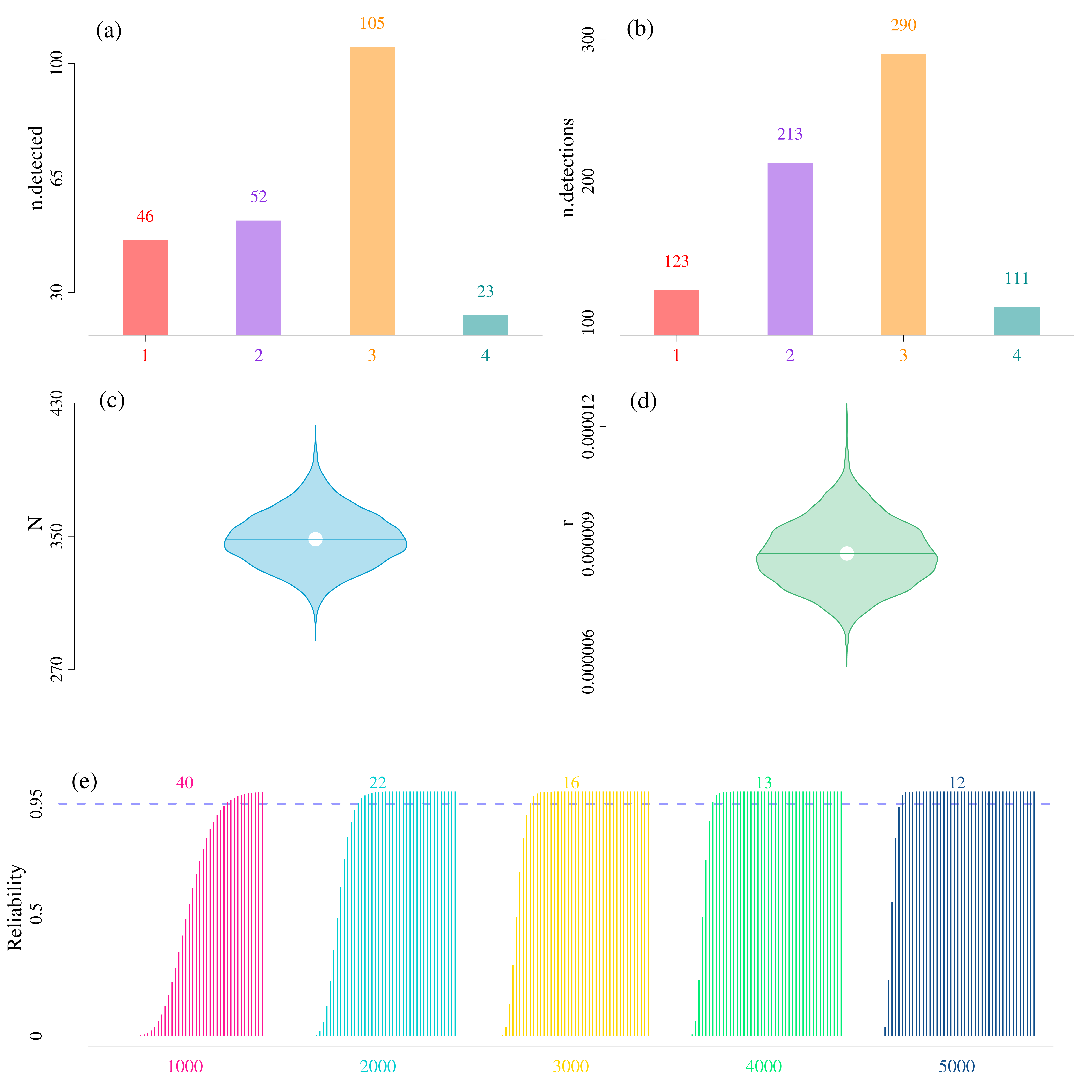}
	\end{tabular}}
	\caption{\footnotesize Details of commercial software testing data summary and parameter estimates. Panels (a) and (b): comparison of number of detected bugs (panel a) and total number of detections across all the bugs, i.e., $\sum_{i=1}^M\sum_{j=1}^J y_{ij}$ (panel b) in each phase of software testing data set. Panels (c) and (d): Posterior density violins of the population size estimator $N$ (panel c) and detection parameter $r$ (panel d) for each simulation scenarios. Panel (e): Estimates of posterior reliability with threshold 100 and stopping phase for attaining optimum reliability level 0.95. The bars represent posterior reliability at phases 1,2,\dots,50. Each barplot in panel (e) corresponds to a distinct number of test inputs (given along the $x$-axis) that was used in each future phases.}
	\label{fig:SoftwareBugData.results}
\end{figure*}

The posterior estimates of the main parameters $N$, $\psi$, $r$ and $B_4$ are provided in Table~\ref{table:Bug.est} and visually portrayed in Figure~\ref{fig:SoftwareBugData.results}. The posterior mean estimate of the total number of bugs was 348 with a 95\% credible interval (317, 382). The posterior mean of inclusion probability $\psi$ was estimated at 0.696 with a 95\% credible interval (0.618, 0.774). The estimate of $\psi$ also confirmed that the upper bound $M=500$  we had set was sufficiently large enough to not to influence in the estimation of $N$. Although the posterior mean estimate of size-biased detection model parameter $r$ was estimated at a very small magnitude $8.761 \times 10^{-6}$, we had coded the parameter with a logistic transformation to retain the accuracy in estimation and MCMC mixing. The remaining eventual bug size after the 4 testing phases was estimated as 703 with a 95\% credible interval (457, 1006). Here we have assumed that the number of test cases in each future phase to be 3000 in order to resemble with the observed data set. 



We found the reliability to attain the target 95\% level at phase 16 if we would have continued with 3000 test cases in each phase, implying the developer would need to continue software testing for 12 more future phases (after the 4 testing phases already conducted) to attain the targeted software reliability level. Hence, the stopping phase was estimated as 16. The reliability took much longer (40 phases) to reach the targeted 95\% level with 1000 test cases in each phase, and took only 12 phases with 5000 test cases in each phase (these results are provided in the appendix). This also revealed that it takes approximately 36000 future test cases to attain the targeted reliability of 95\%.    


\end{multicols}
\begin{table*}[ht] \centering 
	\caption{\footnotesize Estimates of different parameters in the data analysis of commercial software testing data (Section~\ref{sec:application.software.bugdata}).} 
	\label{table:Bug.est} 
	{\footnotesize
	\begin{tabular}{
	@{\extracolsep{10pt}} l  @{\extracolsep{20pt}}  r @{\extracolsep{20pt}}  r @{\extracolsep{20pt}}  r @{\extracolsep{20pt}} r } 
		\\[-1.8ex]\hline 
		\hline \\[-1.8ex] 
Parameter  & Mean & SD & 2.5\% & 97.5\%  \\ 
 &  &  & Quantile & Quantile  \\ 
		[0.5em] \hline \\[-1ex]  
$N$ & 348 & 17 & 317 & 382 \\ 
		$\psi$ & 0.696  & 0.040  &  0.618  &  0.774 \\ 
$r$  & $8.761 \times 10^{-6}$  & $8.331 \times 10^{-7}$ &  $7.261 \times 10^{-6}$  &  $1.044 \times 10^{-5}$  \\ 
	$B_4$ & 703 & 141 & 457 & 1006 \\

		\hline \\[-1.8ex] 
	\end{tabular} }
\end{table*} 


\begin{table*}[ht] \centering 
	\caption{\footnotesize Estimates of different parameters in the ISRO mission software testing data analysis (Section~\ref{sec:application.ISRO.data}).} 
	\label{table:ISRO.est} 
	{\footnotesize
		\begin{tabular}{
				@{\extracolsep{10pt}} l  @{\extracolsep{20pt}}  r @{\extracolsep{20pt}}  r @{\extracolsep{20pt}}  r @{\extracolsep{20pt}} r } 
			\\[-1.8ex]\hline 
			\hline \\[-1.8ex] 
			Parameter  & Mean & SD & 2.5\% & 97.5\%  \\ 
			&  &  & Quantile & Quantile  \\ 
			[0.5em] \hline \\[-1ex]  
			
			$N_G$ & 84 & 2 & 80.000 & 89.000\\ 
			$\psi$ & 0.257  & 0.032  &  0.195  &  0.323 \\ 
			$N$ & 94 & 1 & 94.000 & 95.000 \\ 
			$r$  & $1.102 \times 10^{-3}$  & $3.006 \times 10^{-4}$ &  $6.439 \times 10^{-4}$  &  $1.807 \times 10^{-3}$ \\ 
			
			\hline \\[-1.8ex] 
	\end{tabular} }
\end{table*}




\begin{multicols}{2}

\section{Application to ISRO mission empirical data}\label{sec:application.ISRO.data}

\subsection{Data description}\label{sec:ISRO.data}

The ISRO data set consists of the outcomes from software testing conducted on each of the 5 softwares during 35 missions. Each of the softwares had been updated before different missions were executed. There were 3 primary stages of software testing: (i) `Code inspection' (CI) where a group of experts manually tests each of these softwares in search of potential bug(s), (ii) `Module testing' (MT) where different parts or modules of these softwares are tested, (iii) `Simulation testing' (ST) where numerous inputs are run through the software in seven different phases, viz., SIP, SFIT, IPT, Stress OILS, HLS, ALS and Performance OILS. Different number of bugs were detected during these three primary stages: $ n_{CI} = 33 $ bugs were detected during CI stage, $ n_{MT} = 27 $ bugs were detected during MT and $ n_{ST} = 34 $ bugs were detected during ST (where the phase specific segregation is as the following: $n_1$ = 9, $n_2$ = 7, $n_3$ = 7, $n_4$ = 8, $n_5$ = 1, $n_6$ = 2, $n_7$ = 0). There were also different number of test cases for each mission in each software and in each phase.  For our analysis we consider the testing data from MT and seven phases of ST (i.e., $ Q = 8 $ testing phases) in total as observed data set. We use the detections during CI as deterministic constant because of the lack of probabilistic structure of this testing phase.


\subsection{Results from ISRO mission data analysis}\label{results:isro.data}

We applied the grouped version of our size-biased model (Section~\ref{sec:groupmodel}) to ISRO mission data set which was perfectly suited for applying this model. The different missions, different softwares used in those missions and the different phases - all contributed to the variation of groups and number of bugs in a group. In the observed data set, any change in the mission, software or phase was considered as a different group formation. Here, it is not possible to extend the number of phases, hence instead of finding a stopping phase, we obtain the number of  future test cases required to get the remaining bug size below a pre-specified threshold. This future test cases can be implemented before a future mission or after a software update. 

The posterior mean of number of groups of bugs was estimated at 84 with a 95\% credible interval (80, 89) (see Table~\ref{table:ISRO.est}). The posterior mean estimate of $\psi$ is 0.257 with a 95\% credible interval (0.195, 0.323). This also confirms our specified upper bound $M_G=200$ for the number of groups to be appropriate. The size-biased detection model parameter is estimated as $1.102 \times 10^{-3}$  with a  95\% credible interval  ($6.439 \times 10^{-4}$, $1.807 \times 10^{-3}$). The total number of bugs present was estimated as 94 with 95\% credible interval (94,95) which is highly precise.

\begin{figure*}[ht] 
	\centering
	{\footnotesize \begin{tabular}{@{\hspace{-0.1cm}} l } 
			\includegraphics[scale=0.19]{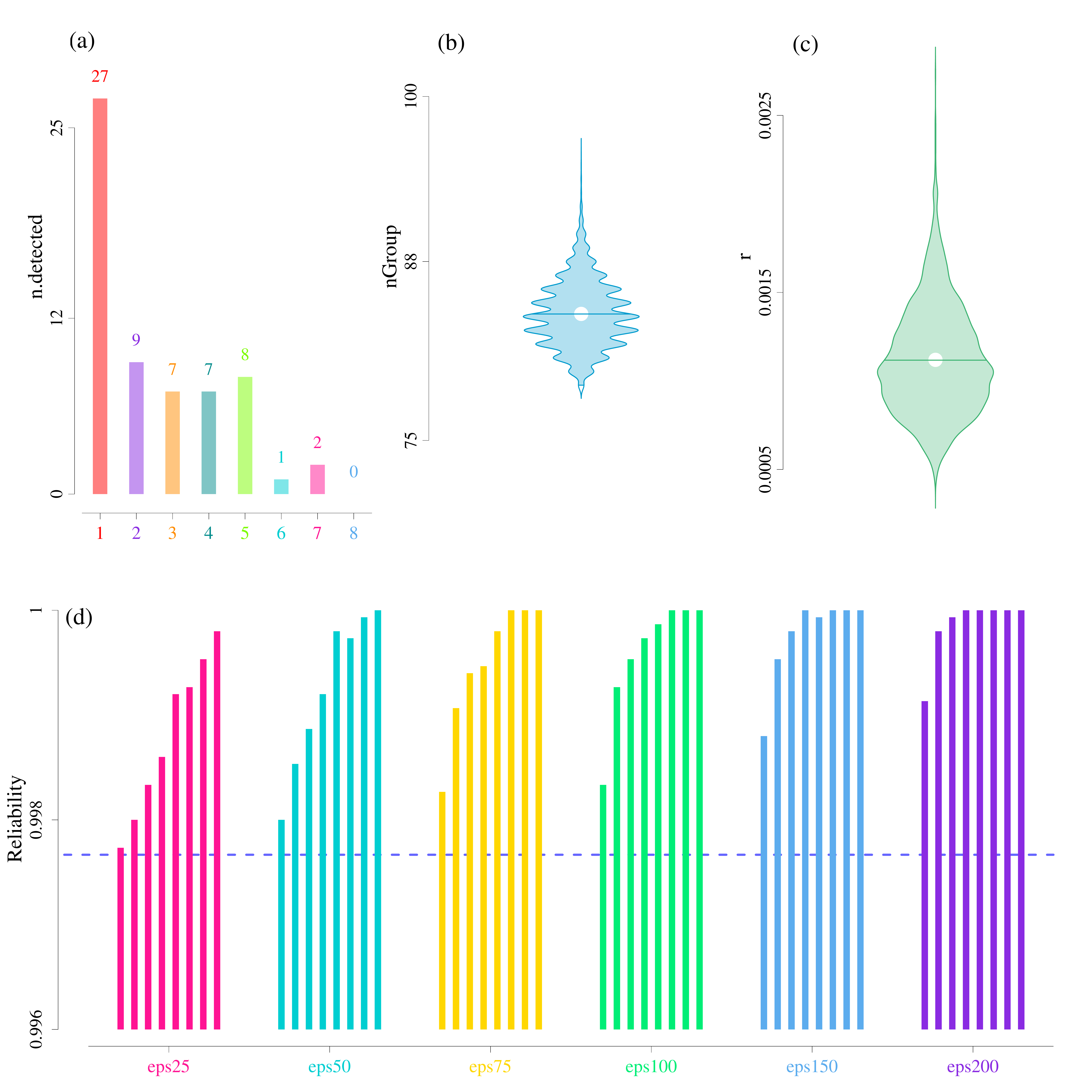}
	\end{tabular}}
	\caption{\footnotesize Details of ISRO mission software testing data summary and parameter estimates. Panel (a): Number of detected bugs (panel a) in each phase of ISRO mission data set. Panels (b) and (c): Posterior density violins of the population size estimator $N$ (panel b) and detection parameter $r$ (panel c). Panel (d): Estimates of posterior reliability with different thresholds 25,50,75,100,150,200. The horizontal dotted line represent the reliability estimate after first 8 testing phases. The bars in each barplot correspond to different numbers of future test cases 25,50,75,100,150,200,250,300. Each barplot in panel (d) corresponds to a distinct threshold (given along the $x$-axis).}
	\label{fig:ISROData.results}
\end{figure*}

The reliability of the softwares is estimated as 0.995 after the 8 testing phases (including module testing and seven phases of simulation testing) with threshold $\eps=25$. Since the testing phases had managed to detect almost all the bugs present in the softwares, this has led to such high reliability. We also show that reliability increases with the increase in number of future test cases (Figure~\ref{fig:ISROData.results}).

\section{Discussion}\label{sec:discussion}

We described a Bayesian generalized linear mixed model that can be applied to software testing detection data set to explicitly model and estimate the population size, detection probability and latent size of the bugs. The model also allows estimation of software reliability for any given threshold (Section~\ref{sec:bugmodel}). Consequently, we could obtain an estimate of the stopping phase providing the number of additional phases of testing are required to achieve an optimum reliability level (say, 0.95). 

We showed via a simulation study that the parameters of interest (e.g., $N$, $r$, reliability) can be accurately estimated by our model. Number of inputs plays a key role in software testing in general, as higher number of inputs boosts the probability of detecting of bugs (Table~\ref{table:ndets}). This also led to more accurate estimation of the model parameters, which can be observed in the lower magnitude of CV estimates of $N$ and $r$ with higher number of inputs  (Table~\ref{table:N.est}). Further, we also noticed that, in such scenarios, threshold reliability level was attained comparatively quicker than the scenarios with lower number of inputs (Figure~\ref{fig:sim.results}e). 

Size biased model fitted to empirical software testing data of bugs yielded satisfactory estimates of the key parameters. However, we noticed that the software testing conducted were rather inefficient since the estimated software reliability was approximately near zero after the first four phases of testing (Figure~\ref{fig:SoftwareBugData.results}). We anticipate that some major bugs (with moderately large size) were still present. We receommend to continue testing for at least 36000-40000 more test cases (which could be broken down into multiple phases) to attain the desired software reliability level 95\%. 

On the contrary, software reliability estimates of ISRO mission softwares were found to be extremely high (i.e., 0.998) after the first 8 testing phases, demonstrating the advantage of efficient software testing. Our finding that the number of bugs detected were almost equal to the true number of bugs available to be detected also supports this.

The developed model can also be used for similar problems in the other fields. For instance, in hydrocarbon exploration, digging a field can be considered analogous with testing a software with different inputs, outcome of which can be considered either as a success (implying sufficient hydrocarbon has been found after digging) or as a failure (implying that the digging did not yield sufficient hydrocarbon which may be viable). 

Given the enormous amount of interest in software testing in technology sector, our size-biased model could be very useful to provide accurate estimates of the number of present bugs as well as software reliability. Our model used the Bayesian paradigm which added the required flexibility to estimate a large number of model parameters. Although we found the parameter estimates to be moderately robust, we recommend to conduct a prior sensitivity study before application of the size-biased model.





\vspace{-0.3cm} 

\section*{Conflicts of interest}
It is hereby declared that the authors do not have any conflict of interest.

\section*{Code availability}
R codes for generating simulated data and data analysis are provided in the online supplementary material and also can be found in GitHub \url{https://github.com/soumenstat89/size_biased}. The two empirical data sets on software testing can be accessed from \url{https://dx.doi.org/10.21227/zm2c-3807}.









\bibliographystyle{abbrvnat}
\bibliography{bibliography}

\end{multicols}


 \end{document}